# Multiscale Theory of Finite Size Bose Systems: Implications for Collective and Single-Particle Excitations

S. Pankavich[a,b], Z. Shreif[a], Y. Chen[c], and P. Ortoleva[a]

[a] Center for Cell and Virus Theory, Department of Chemistry, Indiana University, Bloomington, IN 47405
[b] Department of Mathematics, University of Texas at Arlington, Arlington, TX 76019

[c] Department of Physics, Purdue University, West Lafayette, IN 47907

Contact: ortoleva@indiana.edu (812) 855-2717

# ABSTRACT

Boson droplets (i.e., dense assemblies of bosons at low temperature) are shown to mask a significant amount of single-particle behavior and to manifest collective, droplet-wide excitations. To investigate the balance between single-particle and collective behavior, solutions to the wave equation for a finite size Bose system are constructed in the limit where the ratio $\varepsilon$ of the average nearest-neighbor boson distance to the size of the droplet or the wavelength of density disturbances is small. In this limit, the lowest order wave function varies smoothly across the system, i.e., is devoid of structure on the scale of the average nearest-neighbor distance. The amplitude of short range structure in the wave function is shown to vanish as a power of $\varepsilon$ when the interatomic forces are relatively weak. However, there is residual short range structure that increases with the strength of interatomic forces. While the multiscale approach is applied to boson droplets, the methodology is applicable to any finite size bose system and is shown to be more direct than field theoretic methods. Conclusions for $^4$He nanodroplets are drawn.

# 1. INTRODUCTION

Quantum clusters (QCs) are assemblies such as quantum dots [1-3], superfluid droplets [4-6], fermion droplets [7-8], superconducting particles [9,10], and other structures [11,12]. They involve processes simultaneously acting over multiple scales in space and time from the interaction of individual particles to droplet-wide, collective dynamics. There have been a number of studies presenting theories of their properties, and selected citations are provided above. However, none appear to address the multiscale character of a QC or introduce a framework that takes full advantage of the separation of scales as a way to solve the wave equation.

Recently, it was shown that the wave equation for a fermion quantum nanodroplet can be cast in a form that explicitly manifests its multiscale character [13]. This formulation enables a deeper understanding of the interplay of angstrom and nanometer scale processes underlying the unique behaviors of a fermion QC imposed by the exclusion principle. The objective of the present study is to introduce a novel technique that builds-in characteristics of low temperature boson droplets.

The central concept of our earlier approach [13] is that there are two or more distinct spatial scales of motion for the constituent fermions in a QC. One is the short characteristic length, i.e., the average nearest-neighbor spacing. This short scale is $n^{-1/3}$ for average number density $n$ within the QC. A second characteristic length is the QC diameter. The ratio $\varepsilon$ of the smaller-to-larger of these lengths was introduced to enable a perturbation method for solving the wave equation. While the interaction potential for a condensed QC is large, and thereby cannot serve as the basis of a perturbation analysis, $\varepsilon$ for a QC of $10^3$ to $10^6$ particles is small, say $10^{-1}$ to $10^{-2}$. Thus, $\varepsilon$ presents itself as a natural candidate for the basis of a perturbation theory of QCs that could yield accurate approximations, even for strongly interacting systems.

In a boson QC there are two types of processes to be accounted for. The QC-wide processes are either collective (e.g., rotations, coherent density waves, or shape oscillations) or migrations of particle-like disturbances across the QC (i.e., the coordinated motion of a given particle and a set of others responding to the first). For identical quantum particles the latter quasi-particle excitations are not identifiable with a specific particle. In contrast to these global processes, short-scale ones reflect close encounters of particles related to the interparticle potential. For fermions, the exclusion principle strongly affects these short-scale motions. However, bosons display the opposite tendency, i.e., a "quorum principle". At low temperature, all bosons tend to be in the same state, i.e., at $T = 0$ all bosons are in the single particle-like ground state for weakly interacting systems. While the $N$-atom wave function is observed to have some long range structure, as expressed in the low momentum behavior of neutron scattering experiments, it is stated that only ~10% of the $^4$He atoms are participating in this Bose condensate-like behavior [14-16]. This seems to be a reflection of the strength of the interatomic forces in a $^4$He liquid. In the present work we explore this effect, seeking to show how short scale structure increases in intensity as the strength of the interatomic forces increase. In particular, we seek a description wherein the $N$-atom wave function can be understood as a short lengthscale factor with variability that increases with the strength of the potential, multiplied by a long scale (envelope) factor representing collective modes. We suggest that this is a distinct picture from that wherein it is stated that 10% of the particles are in the Bose condensate (BEC) and 90% are in higher momentum dressed single particle states, i.e. the $N$-atom system cannot be understood in this fashion. All atoms participate in the collective motion and are part of the quasi-particle response.

Though the multiscale analysis is demonstrated for boson droplets, the approach also applies to trapped atomic Bose-Einstein condensates, with typical inter-atomic distances between 0.1 and 1 micron, and system size ranging from 10 to 1000 microns, hence containing between $10^3$-$10^6$ atoms [17,18]. In the atomic Bose condensate, the inter-atomic interaction can be tuned by a Feshbach resonance [19]. The study of strongly-interacting trapped BECs (near a Feshbach resonance) has been a subject of active investigation [17,18, 20-25]. Many properties, such as the condensate fraction and collective mode frequencies, can be modified near the Feshbach resonance [2-,22]. Such strongly, yet still short-ranged interacting Bose condensates offer an interesting "middle ground" between the weakly interacting BEC (well described by mean-field theory) and $^4$He superfluid (with strong and longer range interaction). More recently, trapped BECs with long-range dipolar interactions have also generated much interest [26,27].

In the present study, we attempt to rigorously determine the limiting behavior of the wave function for boson QCs as $\varepsilon \to 0$. The objective is to develop a theory of boson droplets by integrating these notions into a multiscale theory valid for strongly interacting finite bose systems described above.

There is a long history of multiscale analysis for $N$-particle systems [28-39]. Most relevant is the analysis of the classical Liouville equation wherein one identifies order parameters that characterize the slow/long lengthscale behaviors of a system, and mesoscopic equations for the stochastic dynamics of order parameters are derived [40-53]. These order parameters describe nanometer scale features such as the position, orientation, and major substructures of a nanoparticle [52-55]. The ansatz starting the analysis is that solutions to the

Liouville equation (i.e., the *N*-particle probability density) depend on the atomistic configuration both directly and indirectly, the latter through the order parameters. Other approaches for quantum systems (e.g., truncated Laplace transformation of the interaction potential and QM/MM [56-65]) have been developed, but are distinct from the $\varepsilon$-length ratio perturbation approach developed here and do not yield a rigorous mesoscopic wave equation for QC dynamics.

The ansatz that starts our multiscale QC analysis is that the wave function $\Psi$ reflects a dual dependence on the configuration of the *N* bosons in the QC. Let $\underline{r} = \{\vec{r}_1, \vec{r}_2, \cdots \vec{r}_N\}$ be the set of positions of the *N* bosons (assumed identical). By choice of units, the position $\vec{r}_i$ of particle *i* is displaced a distance of about one unit when it moves a distance $n^{-1/3}$. In contrast, $\vec{R}_i \, (= \varepsilon \vec{r}_i)$ changes by a distance of about one unit as particle *i* traverses the entire QC. Denote the collection of these scaled positions by $\underline{R} = \{\vec{R}_1, \vec{R}_2, \cdots \vec{R}_N\}$. To capture the distinct types of behavior (long and short scale), we hypothesize the wave function $\Psi$ has a dual dependence on configuration, i.e., $\Psi(\underline{r}, \underline{R})$. This dual dependence is not a violation of the number of degrees of freedom (3*N*). Rather, it is a way to express the distinct ways in which $\Psi$ depends on droplet configuration. We show that if $\varepsilon$ is small, the two distinct dependencies of $\Psi$ can be constructed via a multiscale perturbation technique. Hence, multiscale analysis naturally reveals the implications of these notions for a boson QC.

## 2. MULTISCALE FORMULATION

The behavior of a low temperature boson QC is now explored via multiscale analysis. We demonstrate how the individuality of the particles is lost, yielding QC-wide cooperative dynamics. The absence of a Fermi level makes it difficult to track the number of particles in a given region of space. Thus, the system lapses into a collective, delocalized bosonic presence without a well-defined sense of the individuality of particles. However, strong interactions between particles, as in liquid $^4$He, are expected to induce short-scale character in the wave function. In this section, we show how delocalization can emerge naturally from a multiscale analysis of the wave equation for a boson QC. We first consider the case of relatively weak interactions and then explore the effect of stronger ones and the inclusion of short-scale structure in the wave function.

### A. Weak interactions, delocalization, and residual short-scale structure

To begin the development, we formulate the wave equation in a manner that reveals the low energy excitations of interest at low $T$. While our formulation facilitates the discovery of the nature of the hypothesized delocalized behavior, self-consistency would preclude the drawing of false conclusions since we begin with the full wave equation. Let $U'$ be the $N$-particle potential, $E'_{ground}$ be the ground state QC energy, and define the deviatoric potential $V' = U' - E'_{ground}$. Introduce the characteristic kinetic energy $\hbar^2/mL^2$ for each of the $N$ bosons, where $m$ is the particle mass and $L$ is the size of the QC. The position of particle $i$ is denoted $L\bar{R}_i$ for dimensionless vector $\bar{R}_i$. In these variables, the dimensionless Hamiltonian $H$ is defined via

$$H = -\frac{1}{2}\underline{\nabla}^2 + V \equiv K + V \tag{1}$$

while the dimensional Hamiltonian is $H' = \hbar^2 H / mL^2$. For this $E'_{ground}$-shifted Hamiltonian, the ground state energy is zero and $V = (U - E_{ground})$. The dimensionless wave equation takes the form $H\Psi = E\Psi$ where $E$ is the dimensionless deviatoric energy.

Let $l$ be $n^{-1/3}$, i.e., $l$ is the typical nearest-neighbor distance within the QC of number density $n$. Then the length scale ratio is given by $\varepsilon = l/L$. With the above definitions, $\vec{R}_i$ changes by a distance of about one unit as particle $i$ traverses the entire QC. In contrast, $\vec{r}_i \equiv \varepsilon^{-1} \vec{R}_i$ changes by a distance of about $\varepsilon^{-1}$ as particle $i$ traverses the entire QC and by about one unit when it traverses one nearest-neighbor distance $l$. Thus, the $\vec{R}_i$ are natural for tracking QC-wide disturbances while the $\vec{r}_i$ are ideal for characterizing close particle-particle encounters. With this, our multiscale ansatz is that $\Psi$ has the dependence $\Psi(\vec{r}_1, \vec{r}_2, \cdots \vec{r}_N; \vec{R}_1, \vec{R}_2, \cdots \vec{R}_N; \varepsilon)$. The dual dependence of $\Psi$ does not constitute a violation of the number (3N) degrees of freedom. Rather, we shall show that it is a reflection of our expectation that $\Psi$ depends on the N-boson configuration in two distinct ways. That both dependencies can be constructed is shown below to be achieved in the small $\varepsilon$ limit. The multiscale wave equation with $\underline{r} = \{\vec{r}_1, \vec{r}_2, \cdots \vec{r}_N\}$ and $\underline{R} = \{\vec{R}_1, \vec{R}_2, \cdots \vec{R}_N\}$ follows from this ansatz and the chain rule:

$$(H_0 + \varepsilon H_1 + \varepsilon^2 H_2)\Psi = \tilde{E}\Psi, \tag{2}$$

where $\tilde{E} = \varepsilon^2 E$ and

$$H_0 = -\frac{1}{2}\underline{\nabla}_0^2, \quad H_1 = -\underline{\nabla}_0 \cdot \underline{\nabla}_1, \quad H_2 = -\frac{1}{2}\underline{\nabla}_1^2 + V. \tag{3}$$

Here $\underline{\nabla}_0$ is the $\underline{r}$-gradient and $\underline{\nabla}_1$ is the $\underline{R}$-gradient.

The objective of our multiscale development is to construct an equation for the mesoscopic wave function $\Phi(\underline{R})$ which varies smoothly across the QC:

$$\Phi(\underline{R}) = \int d^{3N}r A(\underline{R} - \varepsilon \underline{r}) \Psi(\underline{r}). \tag{4}$$

The "sampling function" $A$ is a gaussian-like expression which, in $\underline{r}$ space, is centered about $\varepsilon^{-1}\underline{R}$, is unit normalized, and has a half-width that is much greater than one $\underline{r}$ distance, but is much less than the QC diameter. A central theme of this study is that one may derive a self-consistent equation that is closed in $\Phi$ in the small $\varepsilon$ limit.

A perturbation solution to the multiscale wave equation is constructed as a Taylor expansion in $\varepsilon$, i.e, $\Psi = \sum_{n=0}^{\infty} \Psi_n \varepsilon^n$. To $O(\varepsilon^0)$ one obtains the eigenvalue problem with zero potential:

$$H_0 \Psi_0 = \tilde{E}_0 \Psi_0. \tag{5}$$

Since we seek normalizable solutions which decay to zero as $|\underline{r}| \to \infty$, $\Psi_0$ must be independent of $\underline{r}$. This is consistent with the physical nature of the problem, i.e., as a QC is of finite size (about one unit in $\bar{R}_i$) and $H_0$ is a free particle-like Hamiltonian, the lowest order problem only admits an $\underline{r}$-independent solution. The absence of small-scale structure in $\Psi_0$ implies $\tilde{E}_0$ is zero and

$$\Psi_0 = \Phi_0(\underline{R}), \tag{6}$$

for $\Phi_0$ to be determined in higher order. While it is clear that $\Phi \to \Phi_0$ as $\varepsilon \to 0$, it remains to show that $\Phi_0$ can be constricted in a self-consistent procedure.

Using the $O(\varepsilon^0)$ analysis, i.e., $\tilde{E}_0 = 0$, the $O(\varepsilon)$ equation becomes

$$H_0\Psi_1 + H_1\Psi_0 = \tilde{E}_1\Psi_0. \tag{7}$$

Since $\Psi_0$ is independent of $\underline{r}$, $H_1\Psi_0$ vanishes. Hence, the RHS of (7) is independent of $\underline{r}$ and thus $H_0\Psi_1$ must be independent of $\underline{r}$. Using the normalizability and decay conditions (i.e., $\Psi_1$ vanishes at infinity), we find that $\Psi_1$ is independent of $\underline{r}$ and $\tilde{E}_1 = 0$.

To $O(\varepsilon^2)$ one finds that, since $\Psi_0$ and $\Psi_1$ are independent of $\underline{r}$,

$$H_0\Psi_2 + H_2\Psi_0 = \tilde{E}_2\Psi_0. \tag{8}$$

To arrive at an equation for $\Phi_0$, we (a) multiply both sides of (8) by the sampling function $A$ and integrate over $\underline{r}$; (b) use the divergence theorem and properties of $A$; (c) neglect surface-to-volume terms; and (d) use the fact that $\Phi_0$ does not change appreciably within the sampling volume (i.e., the region wherein $A$ is large). With this, one obtains the mesoscopic wave equation

$$\left[-\frac{1}{2}\nabla_1^2 + \tilde{V}(\underline{R})\right]\Phi_0 = \tilde{E}_2\Phi_0, \tag{9}$$

upon noting that $\Phi \to \Phi_0$ as $\varepsilon \to 0$, and defining $\tilde{V}$ via

$$\tilde{V}(\underline{R}) = \int d^{3N}r\, A(\underline{R} - \varepsilon\underline{r})V(\underline{r}). \tag{10}$$

Even if the bare potential is short range (i.e., independent of $\underline{R}$), $\tilde{V}$ depends on it due to the $\underline{R}$-centered local averaging manifest in the sampling function $A$. Since $E$ is the deviatoric energy, $E = 0$ (i.e., $\tilde{E}_n = 0$ for every $n$) must be an eigenvalue corresponding to the ground state solution of the original problem. Hence, $\tilde{E}_2 = 0$ is an eigenvalue of the mesoscopic equation (9), (10).

While single-particle character in $\Phi_0$ is lost, it was present in the original wave equation. The question arises as to how it was lost. This can be addressed by subtracting (9) from the wave equation (8) to $O(\varepsilon^2)$. One obtains

$$-\frac{1}{2}\underline{\nabla}_0^2 \Psi_2 + [V(\underline{r}) - \tilde{V}(\underline{R})]\Phi_0 = 0. \qquad (11)$$

This is a 3N-dimensional Poisson-like equation with "charge density" equal to $-2(V - \tilde{V})\Phi_0$. Being proportional to $\Phi_0$, the source term is limited to the region within the droplet if (9) supports bound-state solutions. Through $V(\underline{r})$ the "charge density" has short-scale (i.e., individual particle) character with variations over distances of order $n^{-1/3}$. This implies that, although all individual particle character in $\Psi_0$ and $\Psi_1$ is lost, there is residual particle character (i.e., oscillations in the wave function across $\underline{r}$) with amplitude of $O(\varepsilon^2)$. For a 1000 boson QC ($\varepsilon = 10^{-1}$), the single particle character of the wave function is two orders of magnitude smaller than the overall profile as expressed in $\Phi_0$. Since $\Psi_2$ is proportional to $\Phi_0$, and $\Phi_0$ is zero outside the QC, particle-like character is confined to the interior of the QC as expected. We conclude that the multiscale approach to boson QCs constitutes a self-consistent picture and yields insights into the nature of low temperature boson QCs when bound-state solutions to the mesoscopic equation (9) exist.

B. Strong interactions and induced short-scale structure

The above development is now revisited, but for the case where the potential is strong, scaling as $\varepsilon^{-1}$; in particular, we let the potential be $\varepsilon^{-1}V$. With this, $H_1$ is now $-\underline{\nabla}_1 \cdot \underline{\nabla}_0 + V - \tilde{V}$. The rationale for subtracting $\tilde{V}$ is clarified below. We find that $\tilde{V}$ is small when the smoothing volume is appreciable, i.e. averaging the large positive core potential with the weaker long-range attractive tail leads to partial cancellation in $\tilde{V}$ (as demonstrated for helium in Sect. 3). Thus, we assume $\tilde{V}$ is O($\varepsilon$). To preserve the full potential, we put $\tilde{V}^*$ in $H_2$, having denoted $\tilde{V}$ as $\varepsilon \tilde{V}^*$.

With $\underline{\nabla}_0 \Psi_0 = 0$ and the above, (7) becomes

$$H_0 \Psi_1 + (V - \tilde{V})\Psi_0 = \tilde{E}_1 \Psi_0. \tag{12}$$

Multiplying both sides by $A$, integrating over all $\underline{r}$, using integration by parts, and neglecting higher order terms in $\varepsilon$, one obtains $\tilde{E}_1 = 0$. Hence, $\Psi_1$ is the solution of

$$\frac{1}{2}\underline{\nabla}_0^2 \Psi_1 = \left[V(\underline{r}) - \tilde{V}(\underline{R})\right]\Phi_0. \tag{13}$$

Separating variables, one solution to (13) is of the form $\Psi_1 = B(\underline{r}, \underline{R})\Phi_0(\underline{R})$ where $B$ satisfies

$$\frac{1}{2}\underline{\nabla}_0^2 B = V(\underline{r}) - \tilde{V}(\underline{R}). \tag{14}$$

Thus, $\Psi$ is seen to have short-scale structure of amplitude O($\varepsilon$), and not O($\varepsilon^2$) as in the case of weaker interactions (Sect. 2A).

The O($\varepsilon^2$) problem now reads

$$H_0 \Psi_2 - \underline{\nabla}_1 \bullet \left(\Phi_0 \underline{\nabla}_0 B\right) - \frac{1}{2}\underline{\nabla}_1^2 \Phi_0 + \left[B\left(V - \tilde{V}\right) + \tilde{V}^*\right]\Phi_0 = \tilde{E}_2 \Phi_0. \tag{15}$$

Upon multiplying both sides of (15) by $A$, integrating over $\underline{r}$, using the divergence theorem and (14), and neglecting surface-to-volume ratio terms, one obtains a mesoscopic wave equation for $\Phi_0$ similar to (9):

$$\left[-\frac{1}{2}\underline{\nabla}_1^2 + C(\underline{R}) + \tilde{V}^*(\underline{R})\right]\Phi_0 = \tilde{E}_2 \Phi_0 \tag{16}$$

$$C(\underline{R}) = \int d^{3N}r A(\underline{R} - \varepsilon\underline{r}) |\underline{\nabla}_0 B(\underline{r}, \underline{R})|^2 . \tag{17}$$

As seen from (14), $B$ is a response to the fluctuations of the potential difference $V - \tilde{V}$. Its gradient with respect to $\underline{r}$ reflects short-scale structure in the derivative potential. Thus one might expect that $|\underline{\nabla}_0 B|^2$ is a type of kinetic energy contribution that adds to the potential $\tilde{V}^*$ driving the dynamics of $\Phi_0$. In the next section, we consider a different analysis using numerical techniques and a calibrated potential for $^4$He.

# 3. APPLICATION TO $^4$He

To explore the implications of the theory of Sect. 2, we developed computational procedures and applied them to $^4$He. We consider factors affecting the behavior of a $^4$He nanodroplet and the structure of our multiscale approach. These include averaging length, kinetic versus potential energy, the effective wavelength of the bosons, and droplet size.

Let the distance over which a Gaussian-like function $\tau$ is appreciable be denoted $\xi$, and let $\tau$ to be unit-normalized. The smoothing function $A$ of Sect. 2 is taken to be a product of $N$ factors $\tau(\vec{R}_i - \varepsilon \vec{r}_i, \xi)$; $i = 1, 2, \cdots, N$, where is $\tau$ a normalized, Gaussian-like sampling function. The smoothed potential $\mathbb{V}$ for pairwise bare interaction potential $v$ takes the form

$$\mathbb{V}(\underline{R}, \xi) = \sum_{i<j} \tilde{v}(R_{ij}, \xi), \tag{18}$$

where

$$\tilde{v}(R_{ij}, \xi) = \int d^3 r_1 d^3 r_2 \tau(\vec{R}_i - \varepsilon \vec{r}_1, \xi) \tau(\vec{R}_j - \varepsilon \vec{r}_2, \xi) v(r_{12}). \tag{19}$$

From symmetry, the smoothed potential $\tilde{v}$ only depends on the distance $R_{ij}$ and, through $\tau$, the smoothing parameter $\xi$.

To investigate the character of $\tilde{v}$, we chose $v(r)$ to be the Aziz potential [66]. A numerical code was written to evaluate the six-fold integral in (19), taking advantage of symmetry. Profiles of $\tilde{v}(R, \xi)$ for various values of $\xi$ are seen in **Fig. 1**. These profiles show that the smoothing parameter has a drastic effect on the position of the minimum, well depth, and overall shape of the potential. As $\xi$ increases, interparticle distance at the minimum and well depth both increase. Thus, as $\xi$ grows larger, $\tilde{v}$ becomes strictly positive and monotonically decreasing since the repulsive core dominates the attractive tail and causes the well to disappear.

The bare potential $v$ has a short-range repulsive core of radius about $l$ and a long-range attractive tail with range of several $l$. With this pairwise interaction, it is expected that if $\xi$ is smaller than $l$, then $\tilde{v}$ is roughly the same as the bare potential $v$. Hence $\Phi_0$ would have short-scale character, in contradiction to our assumption (i.e., $\Phi_0$ depends on $\underline{R}$, not $\underline{r}$). If $\xi$ is large (i.e., similar to the size of the $^4$He nanodroplet), then $\tilde{v}$ is small, (i.e., for most of the range of integration underlying the averaging in $\tilde{v}$, the values of $\tilde{v}$ are small, and contributions from the repulsive core and the longer-range attractive tail tend to cancel). Thus for large $\xi$, $\tilde{v}$ would not support bound states, and the excitation energy $\tilde{E}_2$ depends strongly on the choice of $\xi$. This implies that a self-consistent procedure must be invoked for choosing $\xi$. For example, one solves the mesoscopic wave equation for $\Phi_0(\underline{R}, \xi)$ with corresponding excitation energy $\tilde{E}_2(\xi)$ and then minimizes $\tilde{E}_2$ with respect to $\xi$. Such a strategy is equivalent to constructing the functional $\tilde{E}_2$ whose minimum over all $\Phi_0$ is the solution to the mesoscopic wave equation, and then minimizing this functional with respect to both $\Phi_0$ and $\xi$.

As the profile of the effective potential changes, so does the energy $\tilde{E}_2$. There are several estimates of kinetic and potential energy to be considered. If $L$ is the diameter of the nanodroplet then $\hbar^2/2mL^2$ is the kinetic energy associated with the longest wavelength bosons. In contrast, the rest energy is $\hbar\omega/2$, where $\omega^2 = k/m$ and $k$ is the second derivative of the bare potential evaluated at its minimum. A relevant potential energy is the well-depth for the bare potential, while another is that for $\tilde{v}$. In Table I, we present values of these energies in addition to information about the potential well for differing values of $\xi$ with nanodroplet size $N = 10^3$ and $10^6$. Due to symmetry of (19), the position of the minimum should change by a factor of $\varepsilon$ as $N$ increases from $10^3$ to $10^6$.

At low temperature, boson nanodroplets display collective behaviors wherein individual particle detail gives way to nanoscale order parameter dynamics. Analysis of the mesoscopic wave equation for boson nanodroplets yields implications for $^4$He droplet dynamics including quantized surface waves (i.e., morphological oscillations). Quantized vortices in thin films [67] suggest that there may be related excitations in $^4$He nanodroplets (**Table I**).

Computations with the present theory involve constructing the order parameter $\varphi(\vec{R})$ profile. This can be accomplished via a variational principle based on a functional whose minimum is attained for $\Phi_0$ expressed as an $N$-fold $\varphi$ product. With this, $\varphi$ satisfies

$$\left[-\frac{1}{2}\underline{\nabla}_1^2 + \upsilon_{\text{eff}}\right]\varphi = E^*\varphi, \qquad (20)$$

$$N\upsilon_{\text{eff}}(\vec{R};\varphi) = \frac{\delta}{\delta\varphi(\vec{R})}\int d^{3N}R\,\tilde{V}(\underline{R})\left|\varphi(\vec{R}_1)\varphi(\vec{R}_N)\cdots\varphi(\vec{R}_N)\right|^2 \qquad (21)$$

for functional derivative $\delta/\delta\varphi(\vec{R})$ and three dimensional $\vec{R}$-space Laplacian $\underline{\nabla}_1^2$. This constitutes a nonlinear eigenvalue problem to determine $\varphi$ and the energy $E^*$. States in the form of an $N$-fold product of a single $\varphi$-functions are not necessarily the lowest energy excitations. Expanding the set of admissible functions could lead to lower energy states. Furthermore, that these order parameters can be imaginary is central for capturing droplet analogues of quantized vortices **(Table I)**.

If one adopts a quasi-particle perspective, then the nanodroplet consists of particles of a broad range of wavelengths. The longer correspond to the droplet diameter; for them all detail of the bare potential is lost, i.e., they experience an effective potential which is small, due to averaging the repulsive core and attractive tail. For shorter-wavelength quasi-particles, details of the bare potential are experienced and short-scale structure is induced by the potential. Choosing $\xi$ greater than $l$ but much less than the droplet diameter yields the wave function $(1+B\varepsilon)\Phi_0$ from (13) which captures the range of elementary excitation wavelengths. If $\xi$ is sufficiently large, but still much less than the nanodroplet diameter, then a mean-field approximation for $\Phi_0$ should suffice, i.e., each boson is interacting with many others, so that an effective field is an accurate description.

## 4. RESULTS AND CONCLUSIONS

A mesoscopic wave equation for boson QCs was derived using a multiscale approach. It was shown that properties of boson QCs can be derived via a multiscale perturbation analysis of the wave equation even when interactions between the bosons are strong. Bosons in a droplet at low *T* were shown to act in a collective manner, losing much of their individual character. Low-lying excited state solutions of the mesoscopic wave equation have profiles without spatial variations on the $n^{-1/3}$ scale, at which one would otherwise expect to reflect the presence of individual bosons. Rather, individual particle features merge via the averaging imposed by the smoothness of the wave function. A smoothed effective *N*-particle interaction potential $\tilde{V}$ emerges that only depends on droplet-wide positional information, i.e., not on any $n^{-1/3}$ scale features. With the exception of highly excited states of the droplet, $\tilde{V}$ only supports coherent droplet-wide dynamics.

The key technical achievement of the multiscale analysis is the derivation of a mesoscopic wave equation for boson droplet dynamics. The methodology holds for a strongly interacting boson droplet of finite size at low *T*. The mesoscopic wave function $\Phi_0$ depends on the set of *N*-particle positions $\underline{R}$ which are cast in units such that they change a distance of one unit as a boson traverses the entire droplet. The result is

$$\left[ -\frac{1}{2}\underline{\nabla}_1^2 + \tilde{V}(\underline{R}) \right] \Phi_0 = \tilde{E}_2 \Phi_0 \qquad (22)$$

$$\tilde{V}(\underline{R}) = \int d^{3N}r\, A(\underline{R} - \varepsilon\underline{r}) V(\underline{r}) \qquad (23)$$

where $\underline{\nabla}_1^2$ is the 3*N*-dimensional Laplacian with respect to $\underline{R}$ and $\tilde{V}(\underline{R})$ is the bare *N*-particle potential with the location of each particle averaged over a sampling volume containing a statistically significant number of bosons; $\tilde{E}_2$ is the excitation energy. This mesoscopic wave equation is remarkable in that it holds for strong interparticle interaction strength as long as $\varepsilon$, the ratio of the average nearest-neighbor spacing to droplet diameter, is small (i.e., for those with 1000 or more bosons).

As $\Phi_0$ is the lowest order wave function (i.e., $\Psi \to \Phi_0$ as $\varepsilon \to 0$), it satisfies boson particle label exchange symmetry:

$$P_{ij}\Phi_0 = +\Phi_0 \qquad (24)$$

for permutation operator $P_{ij}$. As $\Phi_0$ is governed by the smoothed potential $\tilde{V}$, much of the individual particle-particle short-range correlation is diminished. This is essentially a self-consistency argument, i.e., $\Phi_0 \leftrightarrow \tilde{V}$. Particle exchange symmetry and the averaging in $\tilde{V}$ suggest that any one boson is not interacting with particular others (i.e., in contrast to two body interaction). This suggests that to good approximation $\Phi_0 = \varphi(\vec{R}_1)\varphi(\vec{R}_2)\cdots\varphi(\vec{R}_N)$ for order parameter $\varphi(\vec{R})$. The single particle density $\rho(\vec{R})$ is defined via

$$\rho(\vec{R}) = \int d^{3N}R_i \sum_{i=1}^{N} \delta(\vec{R} - \vec{R}_i)|\Phi_0|^2, \qquad (25)$$

and hence is approximately $N|\varphi(\vec{R})|^2$. Thus, $\varphi$ is directly related to the number density.

The above implies the boson droplet at low $T$ is characterized by the profile of the order parameter $\varphi(\vec{R})$ which is devoid of short (i.e, $n^{-1/3}$) scale features. In this way, all individual particle behavior is lost as $\varepsilon \to 0$. Since the energy for short-range forces (as for $^4$He) determined by (1) is proportional to $N$, the difference in energy for the $N$+1 and the $N$ particle droplet (the chemical potential) is independent of $N$. In this way there is no energy measure of the number of bosons in the droplet. This is in sharp contrast to the situation for fermions (i.e., due to the Fermi level).

The averaging in $\tilde{V}$ suggests it is single particle-like, i.e., each boson evolves in a local potential field:

$$\tilde{V} = \sum_{i=1}^{N} \tilde{v}(\vec{R}_i). \qquad (26)$$

For $^4$He there is an attractive tail in the bare two-body potential. Thus, we expect $\tilde{v}$ to be qualitatively as in **Fig. 1** for a spherical droplet. More generally, the three-dimensional spatial profile of $\tilde{v}$ depends on droplet morphology, and hence on $\varphi$ itself. This is accounted for in the local averaging embedded in $\tilde{V}$. Thus, the order parameter $\varphi$ is determined by a mesoscopic wave equation with mean-field character. The development of the present multiscale approach suggests that this is not just a crude approximation. Rather, it appears to be a consequence of (a) the smallness of $\varepsilon$ for a droplet; (b) the exchange symmetry constraints (the quorum principle) for bosons; and (c) the smooth profile of the wave function for low temperature droplets (i.e., that there is no $n^{-1/3}$ spatial scale structure in $\Psi$ except as an $O(\varepsilon^2)$ correction).

A promising area for future developments is to extend these results to account for the scattering of atoms or molecules from a QC or the effects of external fields. In work in progress we are applying the multiscale approach to trapped BEC systems, as the interaction progresses from weak to strong, and short-range to long-range, to investigate the effects of interaction and to understand the evolution from weakly-interacting BEC to such a strongly interacting Bose system as $^4$He. Such studies could facilitate the design of experiments to validate predictions of the multiscale theory. Other natural extensions include the analysis of more complex droplet such as those composed of fermions and bosons (i.e., $^3$He/$^4$He mixtures), or those with embedded solid nanoparticles or macromolecules.

## ACKNOWLEDGEMENTS

This project was supported in part by the Indiana University College of Arts and Sciences through the Center for Cell and Virus Theory. Additional support was provided by the Lilly Endowment, Inc.

# FIGURES

FIG.1 Graphs of the effective potential for $N = 10^3$ particles in nanodroplet with differing values of smoothing parameter, (a) $\xi = 0.35$; (b) $\xi = 0.60$, (c) $\xi = 0.90$, respectively.

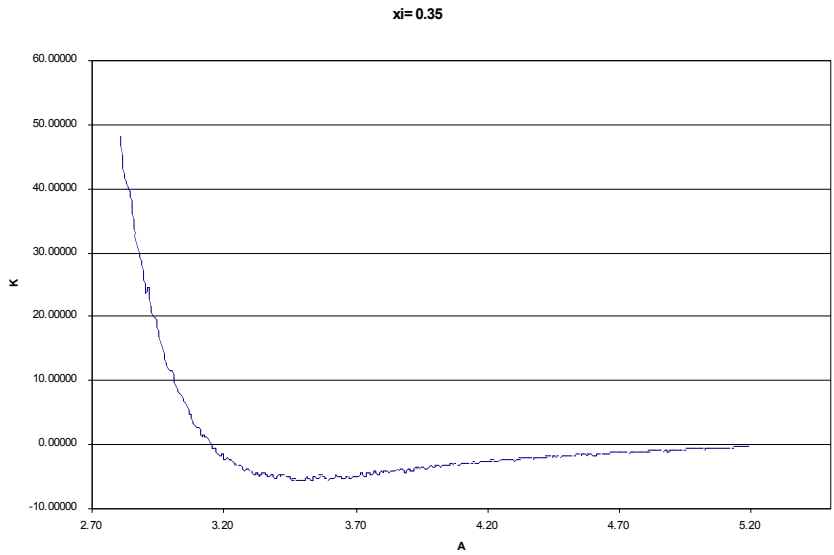

(a)

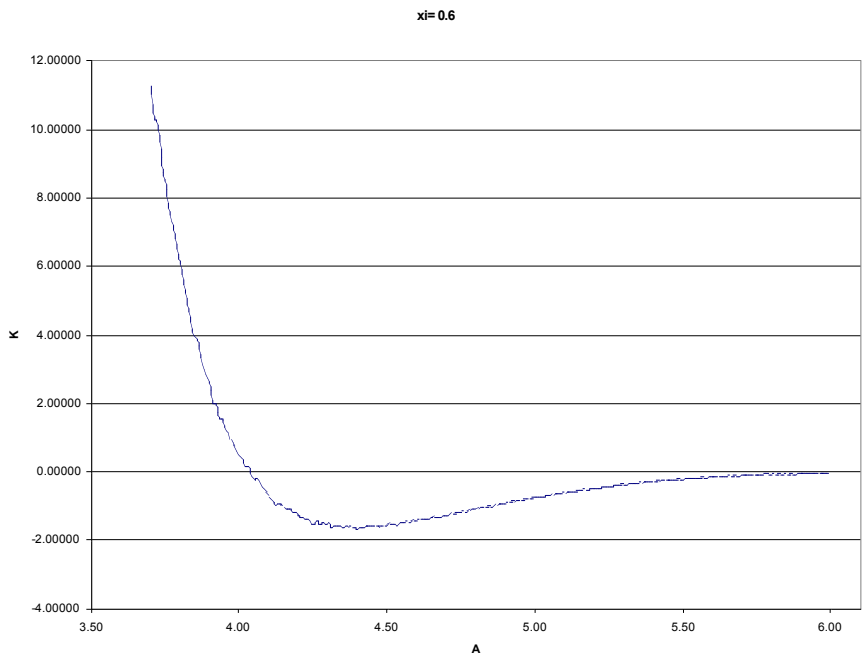

(b)

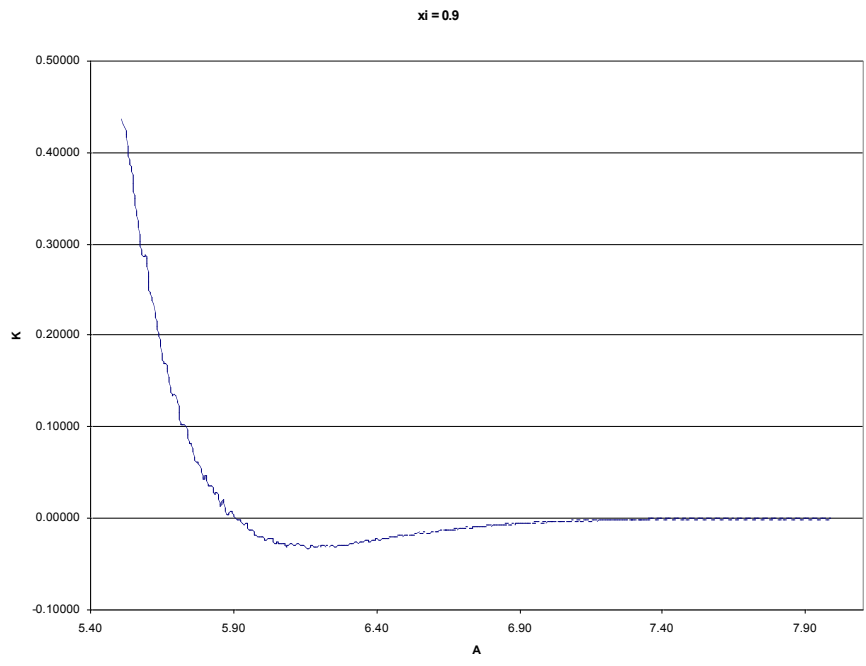

(c)

FIG.2 Schematic depiction of a boson nanodroplet in a vortex-like state of quantized, undamped circulation.

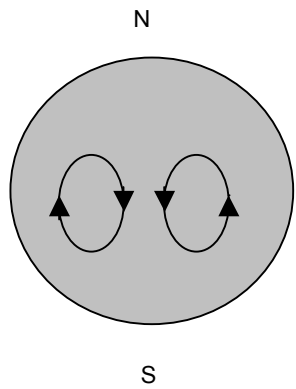

**Table I** Table of values describing energies of the effective (for given value of $\xi$) and bare potentials.

| Size of nanodroplet $N$ | Smoothing parameter $\xi$ | Position of minimum (in Å) | Depth of minimum (in K) | Depth of minimum (in J) | Bare rest energy $k$ (in J) | Harmonic osciallator half-width $L$ (in Å) | Ground state energy $E_0$ (in J) | Kinetic Energy of Nanodroplet (in J) |
|---|---|---|---|---|---|---|---|---|
| | $\xi = 0$ | 2.96 | -10.8 | $-1.49*10^{-22}$ | $1.108*10^{-22}$ | $1.303226*10^{-5}$ | $6.80811*10^{-33}$ | $2.06*10^{-14}$ |
| $N = 10^3$ | $\xi = 0.35$ | 3.52 | -5.57 | $-7.59*10^{-23}$ | | | | |
| | $\xi = 0.60$ | 4.40 | -1.60 | $-2.21*10^{-23}$ | | | | |
| | $\xi = 0.90$ | 6.23 | -0.03 | $-4.14*10^{-25}$ | | | | |
| $N = 10^6$ | $\xi = 0.35$ | 0.35 | -5.57 | $-7.59*10^{-23}$ | | | | |
| | $\xi = 0.60$ | 0.44 | -1.60 | $-2.21*10^{-23}$ | | | | |
| | $\xi = 0.90$ | 0.62 | -0.03 | $-4.14*10^{-25}$ | | | | |